\documentclass[letterpaper, 10 pt, conference]{ieeeconf}

\IEEEoverridecommandlockouts                              
\usepackage{pgfplots}
\pgfplotsset{compat=newest}
\usepackage{epsfig} 
\usepackage{amsmath} 
\usepackage{amssymb}  
\usepackage{aligned-overset}
\usepackage{psfrag}
\usepackage{xcolor}
\usepackage{units}
\usepackage{multirow}
\usepackage[top=54pt, left=54pt, right=54pt, bottom=54pt, paper=letterpaper]{geometry}
\usepackage{paralist}
\usepackage[noadjust]{cite}
\usepackage{epstopdf}
\usepackage{booktabs}
\usepackage{url}
\usepackage{stmaryrd}
\usepackage{algpseudocode}
\usepackage{algorithm}
\usepackage{tabularx}

\usetikzlibrary{external}
\usepackage{titlesec}
\usepackage{accents}
\newlength{\dhatheight}

\usepackage{cleveref}
\usepackage{tablefootnote}

\usepackage[normalem]{ulem}

\newtheorem{thm}{Theorem}

\newtheorem{lem}[thm]{Lemma}

\newtheorem{assum}{Assumption}

\newcommand{\R}{\mathbb{R}}

\newcommand{\N}{\mathbb{N}} 
\newcommand{\Z}{\mathbb{Z}}

\newcommand{\Ac}{\mathcal{A}}
\newcommand{\Bc}{\mathcal{B}}

\newcommand{\ct}{\mathtt{ct}}
\newcommand{\sk}{\mathtt{sk}}

\newcommand{\Dec}{\mathsf{Dec}}

\usepackage{MnSymbol}
\DeclareSymbolFont{MNlargesymbols}{OMX}{MnSymbolE}{m}{n}
\makeatletter
\Decl@Mn@Open {\lsem}{MNlargesymbols}{'102}
\Decl@Mn@Close{\rsem}{MNlargesymbols}{'107}
\makeatother

\newcommand{\blue}[1]{#1}

\renewcommand{\mathbf}[1]{\mathrm{#1}}

\def\tvdots{\vbox{\baselineskip=2pt \lineskiplimit=0pt \kern6pt \hbox{.}\hbox{.}\hbox{.}}} 

\title{\bf 
Encrypted system identification as-a-service\\ via reliable encrypted matrix inversion}

\author{%
Janis Adamek$^\ast$, %
Philipp Binfet$^\ast$, %
Nils Schl\"uter, %
and Moritz Schulze Darup%
\thanks{Janis Adamek, Philipp Binfet, Nils Schl\"uter, and Moritz Schulze Darup are with the Control and~Cyber-physical Systems Group, Department of Mechanical Engineering, TU Dortmund University, Germany. E-mail correspondence to  \{janis.adamek, philipp.binfet, nils.schlueter, moritz.schulzedarup\}@tu-dortmund.de.}%
\thanks{$^\ast$ The first two authors contributed equally to this work.}%
\thanks{Financial support by the German Research Foundation (DFG) under the grants SCHU 2940/4-1 and SCHU 2940/5-1 is gratefully acknowledged.}
}

\begin{document}
\maketitle
\thispagestyle{empty}
\pagestyle{empty}

\begin{abstract}
Encrypted computation opens up promising avenues across a plethora of application domains, including machine learning, health-care, finance, and control. Arithmetic homomorphic encryption, in particular, is a natural fit for cloud-based computational services. However, computations are essentially limited to polynomial circuits, while comparisons, transcendental functions, and iterative algorithms are notoriously hard to realize.

Against this background, the paper presents an encrypted system identification 
\blue{service}
enabled by a reliable encrypted solution to least squares problems.
More precisely, we
\blue{devise an iterative algorithm for matrix inversion}
and present reliable initializations as well as certificates for the achieved accuracy without compromising the privacy of provided I/O-data. The effectiveness of the approach is illustrated with three popular identification tasks.

\end{abstract}


\noindent


\section{Introduction}
   In \blue{today's} increasingly connected world, 
   \blue{security is crucial for the safe and reliable operation of} cyber-physical systems.
\blue{Encrypted control (see~\cite{darup2021encrypted} for an introduction) addresses this need by allowing the design and evaluation} of controllers without exposing sensitive information such as system dynamics or input/output signals to (semi-trusted) external parties \blue{(see, e.g., \cite{kim2016encrypting} resp.~\cite{schluter2022encrypted})}.
\blue{In this manner, private control as-a-service is enabled.}
   However,
   many of the existing approaches assume an accurate system model,
   which has to be identified beforehand. 
   \blue{To complete such services, it is desirable to also} privately outsource the system identification process, as knowledge and hardware limitations apply \blue{equally to both} system identification as well as controller design and execution.
   To address this need, we propose an encrypted system identification service.

   \textbf{Related works \blue{and our contributions}.\quad} 
   Many system identification tasks can be reformulated as least squares problems, which is also a classical problem in statistical analysis and machine learning. Therefore, different confidential solution strategies for linear regression have been developed in the emerging field of privacy-preserving machine learning, despite the large focus on logistic regression~\cite{chen2018logistic,kim2018secure,han2019logistic}. 
   The existing literature can be classified by
   application scenarios, encryption methods, optimization algorithms, and provided reliability analysis.
   
   A wide variety of the existing works like~\cite{qiu2020privacy} and~\cite{cock2015fast} focuses on linear regression over distributed data, which enables the use of secure multi-party computation~\cite{zhao2019secure} for the privacy-preserving calculations, but does not coincide with the use case of system identification as-a-service.
   Early works in
   \blue{the}
   client-server setting like~\cite{chen2022implementing} use the partially homomorphic Paillier cryptosystem and are forced to evaluate the matrix inverse locally.
\blue{Iterative methods for computing matrix inverses have been preferred over alternative approaches, as they are less complicated to initialize compared to gradient-based methods~\cite{esperanca2017encrypted} and scale better than calculating adjugate matrices~\cite{wu2012using}.
Publications adjacent to ours that are also based on Newton-type or related methods do not consider a reliable encrypted initialization~\cite{lu2016using,cheon2018multi}, or do so in a rather conservative way~\cite{hall2011secure}.
In the very recent work~\cite{Ahn2024}, the authors give convergence guarantees for solving linear regression problems but they rely on plaintext information for the initialization and do not certify the quality of the solution.}

\blue{%
The present paper addresses
the above shortcomings by providing novel certificates \blue{for} whether a prescribed error bound can be met, while being largely independent of side-knowledge.
Furthermore, \blue{we improve on existing approaches 
by integrating encrypted matrix inversion as a subtask with a more comprehensive and practically relevant privacy-preserving system identification framework via a tailored least squares algorithm}, 
where end-to-end privacy is provided through leveled homomorphic encryption.}

   \textbf{Outline.\quad}
   The remaining paper is structured as follows.
   In Section~\ref{sec:preliminaries}, we will revisit basics on linear system identification and homomorphic encryption and introduce iterative techniques for finding inverses of matrices and scalars.
   \blue{Our} encrypted least squares algorithm \blue{is outlined} in Section~\ref{sec:encrypted-sysid}, where we also give details on a suitable initialization and certificates for the accuracy of our solution.
   \blue{The implementation is discussed in \Cref{sec:implementation-details}} before
   \blue{presenting}
   numerical case studies in \Cref{sec:Numerical} verifying the effectiveness of our proposed method in several settings.

\section{Preliminaries}
\label{sec:preliminaries}

\subsection{Basics on linear system identification}
\label{subsec:sysID}

One of the simplest tasks in the framework of system identification aims for fitting the coefficients $a_i$ and $b_j$ of a single-input single-output (SISO) transfer function
\begin{equation*}
\label{eq:SISO_TF}
  G(z)=\frac{b_m z^m + \dots + b_1 z + b_0}{z^n+a_{n-1} z^{n-1} + \dots + a_1 z + a_0}  
\end{equation*}
to $L$ given I/O samples
\begin{equation*}
U_L:=\begin{pmatrix}
    u(0) \\
    \vdots \\
    u(L-1)
\end{pmatrix} \qquad \text{and} \blue{\qquad Y_L:=\begin{pmatrix}
    \hat{y}(0) \\
    \vdots \\
    \hat{y}(L-1)
\end{pmatrix}},
\end{equation*}
\blue{where $\hat{y}(k)$ reflects the actual system outputs $y(k)$ affected by additive (and independent) Gaussian measurement noise.}
\blue{A standard approach to estimate suitable model parameters (see, e.g.,~\cite{ljung1998system}) is to minimize}
the Euclidean norm of
\begin{align*}
&\begin{pmatrix}
 -\hat{y}(0) & \dots &  -\hat{y}(n-1)   \\
    \vdots & & \vdots \\
      -\hat{y}(\ell-1) & \dots & -\hat{y}(L-2) 
\end{pmatrix}
\begin{pmatrix}
    \hat{a}_{0}  \\
    \vdots \\
    \hat{a}_{n-1}
\end{pmatrix}\\
&+\begin{pmatrix}
    u(0) & \dots & u(m) \\
    \vdots & & \vdots \\
 u(\ell-1) & \dots & u(\ell+m-1) 
\end{pmatrix}
\begin{pmatrix}
    \hat{b}_{0}  \\
    \vdots \\
    \hat{b}_m
\end{pmatrix} - \begin{pmatrix}
    \hat{y}(n)  \\
    \vdots \\
    \hat{y}(L-1)
\end{pmatrix},
\end{align*}
where $\blue{\ell:=L-n}$. 

Similarly, the identification of multi-input multi-output (MIMO) state space models
\begin{align}
\label{eq:state_space}
 x(k+1) &= A x(k) + B u(k) \\
 y(k)&= C x(k) + D u(k) \nonumber
\end{align}
is trivial under the simplifying assumption $C=I$ and ${D=0}$. 
In fact, \blue{additionally assuming that the measurement noise is identically distributed across all outputs\footnote{\blue{Note that more complex covariance models can be considered by performing weighted minimizations in~\eqref{eq:estimateAB} and \eqref{eq:estimateABcal}.}},} \blue{useful} estimates $\hat{A}$ and $\hat{B}$ then follow from minimizing the Frobenius~norm~of
\begin{align}
\nonumber
&\hat{A} \begin{pmatrix}
    \hat{y}(0) & \dots & \hat{y}(L-2) 
\end{pmatrix} + \hat{B} \begin{pmatrix}
    u(0) & \dots & u(L-2) 
\end{pmatrix} \\
\label{eq:estimateAB}
&- \begin{pmatrix}
    \hat{y}(1) & \dots & \hat{y}(L-1) 
\end{pmatrix}.
\end{align}
Another straightforward identification is possible for multi-step subspace predictors of the form
\begin{equation*}
Y_N = \Ac \,\xi + \Bc U_N,
\end{equation*}
where $Y_N$ and $U_N$ stand for I/O sequences of length ${N<L}$ and where $\xi$ refers to a generalized initial condition reflecting, e.g., the past $n$ inputs and outputs, i.e.,
\begin{equation*}
\xi(k):=\begin{pmatrix}
    u(k-1) \\
    \vdots \\
    u(k-n) \\
    y(k-1) \\
    \vdots \\
    y(k-n)
\end{pmatrix}.
\end{equation*}
Indeed, \blue{meaningful estimates $\hat{\Ac}$ and $\hat{\Bc}$} then follow from minimizing the Frobenius norm of

\begin{align}
\nonumber
&\hat{\Ac} \begin{pmatrix}
 \hat{\xi}(n)  & \dots & \hat{\xi}(L-N) 
\end{pmatrix} \\ 
\nonumber
&+\hat{\Bc} \begin{pmatrix}
    u(n)  &  \dots & u(L-N) \\
    \vdots & & \vdots  \\
    u(N-1+n) & \dots  & u(L-1)
\end{pmatrix} \\
\label{eq:estimateABcal}
&- \begin{pmatrix}
    \hat{y}(n)  &  \dots & \hat{y}(L-N) \\
    \vdots & & \vdots  \\
    \hat{y}(N-1+n) & \dots  & \hat{y}(L-1)
\end{pmatrix}.
\end{align}
All three examples of system identification lead to a least squares problem of the form
\begin{equation}
    \label{eq:leastSquaresProb}
    \min_{Z} \| M Z - V \|_F^2,
\end{equation}
where the decision variable $Z$ and the parameter $V$
may be vectors (as for the SISO transfer function case) or matrices (as in the two other cases). 
In all cases, the entries of $M$ and $V$ reflect entries of $U_L$ or $Y_L$.
To specify the solution of~\eqref{eq:leastSquaresProb}, we make the following assumption throughout the paper.

\begin{assum}
\label{assum:fullColRank}
   The matrix $M$ in~\eqref{eq:leastSquaresProb} has full column rank.
\end{assum}

The optimizer to~\eqref{eq:leastSquaresProb} is then given by
\begin{equation}
    \label{eq:bestFit}
Z^\ast:= M^\dagger V
\end{equation}
with the pseudoinverse
\begin{equation}
    \label{eq:pseudoInverse}
M^\dagger:=(M^\top M)^{-1}M^\top.
\end{equation}

\subsection{Basics on {public-key} homomorphic encryption}
\label{subsec:HE}
Public-key homomorphic encryption (HE) schemes augment the basic encryption-decryption functionality of ordinary cryptosystems by the 
capability
of evaluating arithmetic operations on encrypted values (i.e., without intermediate decryption).
This combination of features
fits nicely with the idea of a service for outsourcing a computational task: the client can securely distribute
its
encrypted input data and public key
to an external computing party (server), while keeping the secret key, $\sk$, private. In this way, it is ensured that the server can indeed process the client's data while keeping its privacy untainted.
In the following, we will write $\ct(z)$ to denote an encrypted representation (ciphertext) of a value $z$, such that $\Dec_{\sk}(\ct(z)) = z$, where $\Dec_{\sk}$ denotes decryption \blue{using the secret key}. If $z$ is a matrix, we consider $\ct(z)$ to be the matrix that contains as its elements ciphertexts of the corresponding elements of $z$.

Regarding the computational capability, we require a leveled fully homomorphic 
scheme meaning that a fixed amount of successive encrypted additions and multiplications of the form
\begin{subequations}
\label{eq:he-ops}
\begin{align}
\label{eq:he-cadd}
\Dec_\sk\left(\ct(z_1) \oplus \ct(z_2)\right) &= z_1 + z_2,\\
\label{eq:he-cmul}
\Dec_\sk\left(\ct(z_1) \odot \ct(z_2)\right)  &= z_1 z_2, 
\end{align}
\end{subequations}
 are supported.
More specifically, we will make use of the
state-of-the-art
CKKS cryptosystem~\cite{cheon2017homomorphic}, which is based on the ring learning with errors problem.
In CKKS,
a complex number (message) is encoded into an element of the polynomial ring $\Z_{Q_c}[X]/(X^{N_r}+1)$ prior to encryption, where $Q_c$ and $N_r$ are the ciphertext modulus and ring dimension, respectively.
Notably, the
encoding involves a discretization step, which
implies a limited fixed-point precision.

Now, the leveled nature of CKKS essentially 
means that every ciphertext is associated with a level $l_c$ tied to the ciphertext's modulus via $Q_{l_c} = Q_0 s^{l_c}$, where $Q_0$ is the base or decryption modulus and $s$ is the factor between the moduli of two adjacent levels. At the maximum level $l_{\mathrm{max}}$, we have $Q_{l_{\mathrm{max}}}=Q_c$.
By principle, encrypted multiplication introduces a relatively large error into CKKS ciphertexts, which can, however,
be controlled by means of rescaling the resulting ciphertext.
Rescaling effectively reduces the level
while preserving
the represented value with high precision. Practically speaking,
the number of successive multiplications, $k_{\mathrm{mul}}$, that a given encrypted algorithm can perform, starting from a newly encrypted value, is constrained by $l_{\mathrm{max}}$ as an upper bound. This is all the more critical when it comes to iterative algorithms. Thus, we will discuss $k_{\mathrm{mul}}$ in more detail in \Cref{subsec:Encryption}.

In the following, we will assume that encoding and rescaling are applied where it is appropriate without explicitly mentioning these operations for ease of presentation.
Further, we note that all arithmetic operations on scalar ciphertexts used in the following can be reduced to~\eqref{eq:he-ops}.
Analogously, we consider operations on encrypted matrices to reduce to the available encrypted operations on their entries following the elementwise representation discussed above.
In particular, this implies that encrypted forms of subtraction and exponentiation as well as matrix multiplication and transposition can be readily used.
In terms of the level of security provided by the scheme, the most prominently contributing parameters are the base modulus $Q_0$, the maximum supported depth $l_{\mathrm{max}}$, the scaling factor $s$ and the ring dimension $N_r$. Choices for these values are discussed
in \Cref{subsec:Encryption}.
For a more in-depth yet accessible introduction into HE and CKKS in the context of encrypted control, the reader is referred to~\cite{schluter2023brief}.

\subsection{Basics on iterative matrix inversions and scalar divisions}

As apparent from Section~\ref{subsec:sysID}, inverting a (data) matrix is at the heart of many system identification approaches. Aiming for an encrypted system identification, we need to carry out the inversion in an encrypted fashion.
Now, it is easy to see that many standard procedures for inverting matrices in plaintext are not suitable for an encrypted realization. In fact, efficient algorithms like Gauß-Jordan elimination need decisions based
on
the matrix elements, which is not feasible for homomorphically encrypted matrix elements, since only addition and multiplication operations are available. Further algorithms like Cholesky decomposition need a large amount of successive inversions and square roots of the matrix elements, which leads to large multiplicative depths. As a consequence of the statements from subsection \ref{subsec:HE}, we here consider procedures, which require as few divisions as possible. In this context, one promising approach for inverting $M$ is the iteration
\begin{equation}
    \label{eq:NewtonSchulzIter}
W_{k+1} := (2 I - W_k M) W_k
\end{equation}
due to Schulz~\cite{Schulz1933},
which is guaranteed to converge to~\eqref{eq:pseudoInverse}
for a suitable initialization $W_0$ (see below).
Now, a popular choice for the initialization is of the form $W_0:=\alpha M^\top$, where convergence is guaranteed for any 
\begin{equation}
\label{eq:alphaInterval}
    \alpha \in (0, 2/\sigma_{\max}^2(M))
\end{equation} 
with $\sigma_{\max}(M)$ denoting the largest singular value of $M$.
Clearly, computing a suitable initialization in plaintext is straightforward. However, for the desired encrypted system identification, only an encrypted representation of $M$ will be available.
Hence, we will later compute an over-approximation $\mu$ of $\sigma_{\max}^2(M)$ in an encrypted fashion and set $\alpha$ close to $2/\mu$. This will require the evaluation of
an encrypted division. Fortunately, a division of the form $1/\mu$ can be carried out analogously to \eqref{eq:NewtonSchulzIter}. In fact, for any $w_0 \in (0,2/\mu)$, the iteration
\begin{equation}
    \label{eq:NewtoIterScalar}
    w_{k+1}:=( 2 - w_k \mu) w_k
\end{equation}
converges to $1/\mu$.

\section{Reliable encrypted system identification}%
\label{sec:encrypted-sysid}

In the following, we aim for an encrypted system identification as-a-service. More precisely, we consider a client, which sets up a homomorphic cryptosystem and transmits encrypted I/O-data to the service provider (server), which solves~\eqref{eq:bestFit} in an encrypted fashion using the homomorphisms \eqref{eq:he-ops} and an approximation of $M^\dagger$ via~\eqref{eq:NewtonSchulzIter}. 
The server then returns an encrypted approximation of $Z^\ast$ to the client, who
can decrypt and use the identified model parameters~$\hat{Z}$.

A key feature of our approach is that the client can specify a desired error bound $\epsilon>0$ (in plaintext) and that the server chooses a corresponding number of iterations $k_{\text{inv}}$ such that
\begin{equation}
\label{eq:desiredAccuracy}
   \| Z^\ast - \hat{Z} \|_{\max}= \| M^\dagger V - W_{k_{\text{inv}}} V \|_{\max} \leq \epsilon 
\end{equation}
holds under certain assumptions specified below. As the server will not be able to validate these assumptions (due to encrypted data), encrypted certificates will be provided to the client, which is another feature of our approach.

Next, we present a reliable initialization of~\eqref{eq:NewtonSchulzIter} and then specify a sufficient number of iterations $k_{\text{inv}}$ to achieve~\eqref{eq:desiredAccuracy}. Finally, certificates for the underlying assumptions are discussed in Section~\ref{subsec:certificates}.   

\subsection{Reliable initialization and division}

In order to initialize the iterative matrix inversion~\eqref{eq:NewtonSchulzIter} with a suitable $W_0:=\alpha M^\top$, we need to choose an $\alpha$ as in~\eqref{eq:alphaInterval}. Such a choice is non-trivial if the
server
has only access to an encrypted version of $M$. We will see, however, that additionally providing the magnitude
\begin{equation}
    \beta:=\label{eq:boundBeta}
    \left\|\begin{pmatrix}
    U_L \\
    Y_L
\end{pmatrix}\right\|_\infty
\end{equation}
of the I/O-data in an encrypted fashion is sufficient to make a reliable choice. 
More precisely, we assume that the server obtains $\ct(1/\beta^2)$ (along with the encrypted I/O-data).
The server can then exploit the following bound on $\sigma_{\max}^2(M)$.

\begin{lem}
Let $M \in \R^{l \times \nu}$ and let $\beta$ be as in~\eqref{eq:boundBeta}. Assume the entries of $M$ reflect entries of $U_L$ or $Y_L$. Then,
\begin{equation}
    \label{eq:upperBoundsForSigmaMax}
\sigma_{\max}^2(M) < \| M\|_F^2 \leq l \nu \beta^2.
\end{equation}
\end{lem}

\begin{proof}
The first relation in~\eqref{eq:upperBoundsForSigmaMax} reflects a standard result. In fact, we 
have $\sigma_{\max}^2(M)=\lambda_{\max}(M^\top M)$ and 
\[
\| M\|_F^2 = \mathrm{trace}(M^\top M) = \sum_{i=1}^\nu \lambda_i(M^\top M).
\]
Since $M$ has full column rank by Assumption~\ref{assum:fullColRank}, we have $\lambda_i(M^\top M) > 0$ for every $i \in \{1,\dots,\nu\}$, which establishes the first inequality in~\eqref{eq:upperBoundsForSigmaMax}.
We further have
\[
    \| M\|_F^2 = \sum_{i=1}^\nu \sum_{j=1}^l |M_{ij}|^2 \leq l\nu\beta^2,
\]
which provides the second inequality in~\eqref{eq:upperBoundsForSigmaMax}.
\end{proof}

In the following, the leading idea is to use the over-approximation $\mu:=\| M\|_F^2 $ of $\sigma_{\max}^2(M)$ to derive 
a suitable factor $\alpha$ as in~\eqref{eq:alphaInterval} close to $2/\mu$.
Importantly, it is straightforward to compute $\ct(\mu)$ in an encrypted fashion given encrypted $U_L$ and $Y_L$.
However, as division is not a native homomorphic operation, 
we will rely on~\eqref{eq:NewtoIterScalar} to approximate $\ct(1/\mu)$.
To initialize this iteration with a suitable $w_0 \in (0,2/\mu)$, the latter relation in~\eqref{eq:upperBoundsForSigmaMax} is helpful. In fact, we can specify $w_0$ as $\tau /(l \nu) \cdot \ct(1/\beta^2)$ for any $\tau \in (0,2)$
and evaluate the iteration~\eqref{eq:NewtoIterScalar} in an encrypted fashion.
Now, convergence to $1/\mu$ follows from the relative errors 
\begin{equation}
\label{eq:relErrorDiv}
  e_{k}:=1-w_{k} \mu .  
\end{equation}
In fact, the initialization $w_0 \in (0,2/\mu)$ ensures ${|e_0| < 1}$. Moreover, it is easy to see and well-known that the relation $e_{k+1}=e_k^2$
applies for all $k$.
Hence, we find $e_k=e_0^{2^k}$ and consequently $\lim_{k \rightarrow \infty} e_k =0$. Another useful implication is that $e_k \geq 0$ for $k \geq 1$. Clearly, this implies $w_k \leq 1/\mu$ for $k\geq 1$ according to~\eqref{eq:relErrorDiv}. We will exploit this property and consider $k_{\text{div}}\geq 1$ iterations~\eqref{eq:NewtoIterScalar} in the following. It remains to note that $w_{k+1}=(1+e_k)w_k$ implies $w_k>0$ for every $k \geq 1$ given a positive $w_0$ as considered here.

\subsection{Reliable matrix inversion and system identification}

Once $w_{k_\mathrm{div}}$ has been computed (in an encrypted fashion), the server can use it to initialize $W_k:=\alpha M^\top$ with an $\alpha$ satisfying~\eqref{eq:alphaInterval}. To specify this choice, we study the relative error matrix
\begin{equation*}
E_k :=  I - W_k M
\end{equation*}
analogously to \eqref{eq:relErrorDiv}. Now, it is easy to see that $\|E_0\|_2$ is either determined by the smallest or the largest eigenvalue of the matrix $M^\top M$, which is positive definite due to Assumption~\ref{assum:fullColRank}. In fact, we have
\begin{align}
\label{eq:E0norm}
 \|E_0\|_2&=\| I - \alpha M^\top M \|_2 \\
\nonumber
 &= \max \{1 - \alpha \lambda_{\min}(M^\top M),\alpha \lambda_{\max}(M^\top M)-1\}.   
\end{align}
Due to $\lambda_{\min}(M^\top M)>0$ and $\lambda_{\max}(M^\top M)=\sigma_{\max}^2(M)$, we obtain $\|E_0\|_2<1$ for any $\alpha$ as in~\eqref{eq:alphaInterval}. Moreover, the relation $E_{k+1}=E_k^2$ holds (analogously to $e_{k+1}=e_k^2$). As a consequence, convergence of $E_k$ to the zero matrix is guaranteed. Nevertheless, we are eventually interested in a precise system identification in the sense of~\eqref{eq:desiredAccuracy}.
Clearly,
\begin{equation*}
\| (M^\dagger - W_k) V \|_{\max} \leq  \| (M^\dagger - W_k) V \|_2 \leq  \| M^\dagger - W_k\|_2  \|  V \|_2.
\end{equation*}
Thus, we need to investigate the absolute error matrix $F_k:=M^\dagger - W_k$. Now, for an initialization of the considered form $W_0:=\alpha M^\top$, one can show that $F_k=E_k M^\dagger$. Hence,
\begin{equation*}
\| M^\dagger - W_k\|_2 = \| E_k M^\dagger\|_2 \leq \| E_k\|_2 \|  M^\dagger\|_2.
\end{equation*}
In summary, we aim for finding a $k_{\text{inv}}$ such that
\begin{equation}
\label{eq:sufficientForDesiredAccuracy}
   \| E_{k_{\text{inv}}}\|_2 \|  M^\dagger\|_2   \|  V \|_2 \leq \epsilon. 
\end{equation}
Finding such a $k_{\text{inv}}$ is hard without further specifications. In fact, while $\|  V \|_2$ can be easily bounded from above (see Thm.~\ref{thm:kInvBound} below), doing so for $\| E_{k_{\text{inv}}}\|_2 $ and $\|  M^\dagger\|_2$ is non-trivial. To solve this issue, the server will make two assumptions and provide certificates (specified below) to the client for securely verifying their validity.
\blue{\begin{assum}
$\|E_0\|_2 \leq p$ for a predefined $p\in(0,1)$.
\label{assum:errorbound}
\end{assum}}
This
\blue{assumption}
corresponds to an initialization that is contained in the region of convergence and that has at least distance $1-p$ from its boundary. 
A sufficient condition for this assumption to hold is provided in the following lemma.

\begin{lem}
\label{lem:E0p}
    Let $p \in (0,1)$, $k_{\text{div}}\geq 1$, and $\alpha:=(1+p) w_{k_{\text{div}}}$. 
    Then, $\|E_0\|_2 \leq p$ holds if
\begin{equation}
    \label{eq:traceDetCondition}
    \left(\frac{\mu}{\nu-1} \right)^{\nu -1} \frac{1-p}{1+p} \leq   w_{k_{\text{div}}}\det(M^\top M).
\end{equation}
\end{lem}

\begin{proof}
As apparent from~\eqref{eq:E0norm}, we have $\|E_0\|_2 \leq p$ whenever
\begin{equation}
    \label{eq:pConditions}
    1 - \alpha \lambda_{\min}(M^\top M) \leq p \,\,\,\, \text{and} \,\,\,\,  \alpha \lambda_{\max}(M^\top M)-1 \leq p.
\end{equation}
Due to $w_{k_{\text{div}}} \leq 1/\mu <1/\sigma_{\max}^2(M)$ for $k_{\text{div}}\geq 1$, the latter relation in~\eqref{eq:pConditions} holds by the specification of $\alpha$. 
Since the former relation in~\eqref{eq:pConditions} is equivalent to
\begin{equation}
    \label{eq:lambdaMinAlphap}
\lambda_{\min}(M^\top M) \geq \frac{1-p}{\alpha}=\frac{1-p}{(1+p) w_{k_{\text{div}}}}
\end{equation}
and since $\mu=\|M\|_F^2=\mathrm{trace}(M^\top M)$,
\eqref{eq:traceDetCondition} would ensure its validity if 
\begin{equation}
    \label{eq:lowerBoundLambdaMin}
\lambda_{\min}(M^\top M) \geq \left(\frac{\nu-1}{\mathrm{trace}(M^\top M)}\right)^{\nu-1} \det(M^\top M).
\end{equation}
This completes the proof since the right-hand side in~\eqref{eq:lowerBoundLambdaMin} is a known lower 
bound~\cite[Thm.~1]{Merikoski1997} 
on the eigenvalues of a positive definite matrix.
\end{proof}

\blue{\begin{assum}
$\mu \geq q \beta^2$ for a predefined $q>0$.
\label{assum:qbound}
\end{assum}}
Remarkably, $q=1$ is typically a safe choice (except for the rare case that the data \blue{point} determining $\beta$ in~\eqref{eq:boundBeta} is not contained\footnote{Note, for instance, that $y(L-1)$ is not contained in $M$ (but in $V$) for the three system identifications discussed in Section~\ref{subsec:sysID}.} in $M$).
In combination, the two assumptions allow
\blue{for choosing}
a $k_{\text{inv}} \in \N$ such that \eqref{eq:desiredAccuracy} holds as specified next.

\begin{thm}
\label{thm:kInvBound}
Let $p$, $k_{\text{div}}$, and $\alpha$ be as in Lemma~\ref{lem:E0p}. Let $\beta$ be as in~\eqref{eq:boundBeta}, let $q >0$, and let $\epsilon>0$. 
Assume the entries of $M \in \R^{l \times \nu}$ and $V \in \R^{l \times r}$ reflect entries of $U_L$ or $Y_L$. Further assume that $\|E_0\|_2\leq p$, $\mu \geq q \beta^2$, and
  \begin{equation}
  \label{eq:epsilonCondition}
\epsilon < p \sqrt{\frac{1+p}{1-p} \frac{l r}{q}}.
  \end{equation}
Then, \eqref{eq:desiredAccuracy} holds for any $k_{\text{inv}} \in \N$ satisfying
  \begin{equation}
  \label{eq:kInvLowerBound}
      k_{\text{inv}} \geq  \log_2 \left( \frac{\log_2\left( \epsilon \sqrt{\frac{1-p}{1+p} \frac{q}{l r}} \right)}{\log_2(p)} \right). 
  \end{equation}
\end{thm}

\begin{proof}
    We obviously have $\|V\|_2\leq \|V\|_F \leq \sqrt{l r} \beta$ (analogously to~\eqref{eq:upperBoundsForSigmaMax}). Moreover, we find
    \begin{equation*}
    \|M^\dagger\|_2=\frac{1}{\sigma_{\min}(M)} = \frac{1}{\sqrt{\lambda_{\min}(M^\top M)}} \leq \sqrt{\frac{(1+p) w_{k_{\text{div}}}}{1-p}}
    \end{equation*}
    according to~\eqref{eq:lambdaMinAlphap}. Due to $k_{\text{div}}\geq 1$ and $\mu \geq q \beta^2$, we further have $w_{k_{\text{div}}}\leq 1/\mu \leq 1/(q \beta^2)$. Hence,
    \begin{equation*}
     \|M^\dagger\|_2 \leq \frac{1}{\beta}\sqrt{\frac{1+p}{(1-p)q}}.
    \end{equation*}
Now, using $\|E_0\|_2\leq p$ and $E_{k+1}=E_k^2$, we additionally derive
    $\|E_{k_{\text{inv}}}\|_2\leq p^{2^{k_{\text{inv}}}}$. In combination, \eqref{eq:sufficientForDesiredAccuracy} and, thus, \eqref{eq:desiredAccuracy} hold if
    \begin{equation*}
p^{2^{k_{\text{inv}}}} \sqrt{\frac{1+p}{1-p} \frac{l r}{q}} \leq \epsilon.
    \end{equation*}
Deriving a lower bound on $k_{\text{inv}}$ for this condition to hold is straightforward and leads to \eqref{eq:kInvLowerBound}, where \eqref{eq:epsilonCondition} ensures $k_{\text{inv}}>0$ and a well-defined right-hand side in~\eqref{eq:kInvLowerBound}.
\end{proof}

\subsection{Validation via encrypted certificates}

\label{subsec:certificates}

It remains to specify the concept and the realization of the certificates associated with the two assumptions, i.e., $\|E_0\|_2\leq p$ and $\mu \geq q \beta^2$. Given that inequalities are hard to validate using (arithmetic) HE, the core idea is to compute the left- and right-hand side\blue{s} of these (or related) inequalities in an encrypted fashion and to leave their validation to the client. Regarding the second assumption, this is trivial as $\ct(\mu)$ is computed anyway and since $\ct(1/\beta^2)$ is available. Hence, the server can easily compute $\ct(\mu) \odot \ct(1/\beta^2)$ 
and leave the verification of $\mu/\beta^2 \geq q$ to the client. 
Regarding the first assumption, we rely on the sufficient condition~\eqref{eq:traceDetCondition}. 
Here, both sides can be evaluated based on solely encrypted multiplications and additions since $\nu$ and $p$ are given in plaintext and since the determinant can be evaluated, e.g., using Laplace expansion. However, directly evaluating~\eqref{eq:traceDetCondition} typically leads to large left- and right-hand sides, which might result in overflow. Hence, we suggest to downscale both sides by multiplying them with $(1/\beta^2)^\nu$. This leads to the equivalent condition
\begin{equation}
    \label{eq:scaledCertificate}
\left(\frac{\mu}{ \beta^2} \frac{1}{\nu-1} \sqrt[\nu-1]{\frac{1-p}{1+p}} \right)^{\nu -1} \frac{1}{\beta^2} \leq   w_{k_{\text{div}}}\det\left( \frac{M^\top M}{\beta^2} \right),
\end{equation}
which can still be evaluated effectively using HE due to the availability of $\ct(1/\beta^2)$. 
We give some additional details on the implementation of the certificates and the encrypted system identification itself in the following section.

\section{Implementation Details}%
\label{sec:implementation-details}

In the previous section, we provided the theoretical basis for a reliable encrypted system identification. To provide a comprehensive presentation of the resulting service, we now shed light on some implementation details.
To this end, we first provide an overview of the functional blocks and flow of data in \Cref{subsec:Overview}. In \Cref{subsec:iterations-goldschmidt-style}, we discuss some algorithmic improvements to render the encrypted implementation more efficient.
Finally, specifications of the utilized cryptosystem are give in \Cref{subsec:Encryption}.

\subsection{Overview}
\label{subsec:Overview}
\begin{figure}[ht]
  \centering
  {%
  \includegraphics{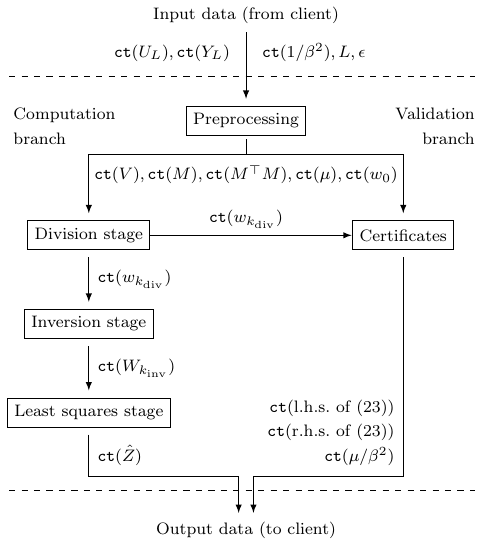}%
  }
  \caption{Overview of the proposed encrypted system identification.}%
  \label{fig:algorithm-overview-high-level}
\end{figure}

\Cref{fig:algorithm-overview-high-level} provides an overview of the functional blocks that make up our system identification service. 
From the client's perspective, the interaction with the server consists in submitting a request with the mandatory data to the server
and eventually receiving the identified system parameters as well as certificates for their accuracy as a response. 
On the sever side, the encrypted system identification builds on the following steps.
During the \textbf{preprocessing}, 
upon receiving $\ct(U_L)$, $\ct(Y_L)$, $\ct(1/\beta^2)$, $L$, and $\epsilon$ from the client, the server assembles $\ct(M)$ and $\ct(V)$. Furthermore, it chooses $p \in (0,1)$ and $q>0$ (typically close to $1$) as well as $\tau \in (0,2)$ (typically close to $2$).
It then computes $\ct(M^\top M)$, $\ct(\mu)$ (while exploiting $\mu=\mathrm{trace}(M^\top M)$), and $\ct(w_0)= {\tau}/{(l\nu)}\cdot\ct(1/\beta^2)$.  Finally, the server selects a $k_\mathrm{div}\geq 1$ and a $k_\mathrm{inv}$ satisfying~\eqref{eq:kInvLowerBound}. After preprocessing, the algorithm splits into the computation and validation branches (see Fig.~\ref{fig:algorithm-overview-high-level}).
The first block in the computation branch is the \textbf{division stage}. Here, $\ct(w_{k_\mathrm{div}})$ is computed by evaluating~\eqref{eq:NewtoIterScalar} in an encrypted fashion based on the initialization $\ct(w_0)$.
Next, during the \textbf{inversion stage}, the server first computes $\ct(\alpha)=(1+p)\cdot\ct(w_{k_\mathrm{div}})$. It then computes
$\ct(W_{k_\mathrm{inv}})$ via \eqref{eq:NewtonSchulzIter} by exploiting that the computation of $\ct(W_{1})$ involves $\ct(M^\top M)$. The computation branch is completed by the \textbf{least squares stage}, where $\ct(\hat{Z}) = \ct(W_{k_\mathrm{inv}})\ct(V)$ is evaluated.
In the validation branch, the server prepares the two \textbf{certificates} specified in Section~\ref{subsec:certificates}. 
The second certificate only requires to compute $\ct(\mu) \odot \ct(1/\beta^2)$. For the first certificate building on~\eqref{eq:scaledCertificate}, the server picks up this result and initially computes
\begin{equation*}
\left( \ct(\mu) \odot \ct(1/\beta^2) \right) \cdot \left(\frac{1}{\nu-1} \sqrt[\nu-1]{\frac{1-p}{1+p}}\right).
\end{equation*}
Raising it to the power of $\nu-1$ is implemented in terms of $\odot$ using binary exponentiation.
The evaluation of the left-hand side in~\eqref{eq:scaledCertificate} is completed by multiplication with $\ct(1/\beta^2)$. Regarding the right-hand side, the server first computes $\ct(M^\top M) \odot \ct(1/\beta^2)$ and then uses Laplace expansion to evaluate the encrypted determinant before multiplying the result with $\ct(w_{k_\mathrm{div}})$. After completing all computations, the server sends $\ct(\hat{Z})$ to the client together with the encrypted left- and right-hand sides of the inequalities~\eqref{eq:scaledCertificate} and ${\mu/\beta^2 \geq q}$.

\subsection{Iteration variants with smaller \blue{multiplicative} depth}%
\label{subsec:iterations-goldschmidt-style}

As already briefly noted in Section~\ref{subsec:HE}, a limiting factor for the encrypted implementation of algorithms is the required multiplicative depth. 
Fortunately, tailored reformulations often allow to reduce the multiplicative depth. 
For instance, a reformulation of the iterations \eqref{eq:NewtonSchulzIter} and \eqref{eq:NewtoIterScalar} here allows to halve the multiplicative depth per iteration.
As a preparation, we note that the multiplicative depth of each original iteration is two. Now, we will see that \eqref{eq:NewtonSchulzIter} can be replaced by the iterations
\begin{subequations}
    \label{eq:GoldschmidtIter}
\begin{align}
    \label{eq:GoldschmidtIterF}
    F_{k+1} &= (2I - H_{k}) F_k \quad \text{and} \\ 
    \label{eq:GoldschmidtIterH}
    \quad H_{k+1} &= (2I - H_{k}) H_k.
\end{align}
\end{subequations}
In fact, 
the following equivalence applies.
\begin{lem}
\label{lem:equivalence-Newton-Goldschmidt}
    Let $F_0:=W_0$ and $H_0:=W_0 M$. Then, the iterations \eqref{eq:NewtonSchulzIter} and \eqref{eq:GoldschmidtIter} are such that $F_k=W_k$ for every~${k \in \N}$.
\end{lem}

\begin{proof}
    We first show that $H_k=W_k M$ applies for every $k \in \N$ by mathematical induction. Clearly, the relation trivially holds for $k=0$ by definition of $H_0$. For the induction step, we subsequently find
    \begin{equation*}
    H_{k+1} = (2I - W_k M) W_k M = W_{k+1} M
    \end{equation*}
    according to~\eqref{eq:GoldschmidtIterH} and \eqref{eq:NewtonSchulzIter}.
    With $H_k=W_k M$ and $F_0=W_0$, it is then  easy to see that \eqref{eq:GoldschmidtIterF} is equivalent to \eqref{eq:NewtonSchulzIter}.
\end{proof}

Now, while being equivalent to~\eqref{eq:NewtonSchulzIter}, the iterations~\eqref{eq:GoldschmidtIter} only have a multiplicative depth of one each. Since they can be carried out in parallel, the multiplicative depth can indeed be halved. Moreover, an analogue approach can be applied to \eqref{eq:NewtoIterScalar}. At this point, it is important to note that similar reformulations have already been used in other works (see~\cite{Ahn2024,hall2011secure}). Furthermore, the iteration~\eqref{eq:GoldschmidtIter} ultimately traces back to Goldschmidt's division algorithm~\cite{Goldschmidt1964}.

\subsection{Specifications of the cryptostem} 
\label{subsec:Encryption}

As mentioned in Section~\ref{subsec:HE}, we implement our scheme using the CKKS cryptosystem as realized in the OpenFHE~\cite{al2022openfhe} library. 
To parametrize the cryptosystem, the required multiplicative depth is again crucial.
To specify this depth, we initially summarize the multiplicative depth required per block in Table~\ref{tab:depthPerStage}. Here, it is important to note that the utilization of the Goldschmidt-style iterations~\eqref{eq:GoldschmidtIter} reduces the required depth per iteration but it increases the effort for the initialization. Nevertheless, we can build upon the preprocessed data when initializing $H_0$ and its scalar counterpart. 
Regarding the certificates, the maximal depth is typically determined by the depth $\nu-1$ required to compute the determinant and the two accompanying multiplications (leading to $\nu+1$ as in Tab.~\ref{tab:depthPerStage}).
However, for small $\nu$, also the depth of the power expression or that underlying $\ct(w_{k_{\text{div}}})$ can dominate. Hence, we formally find $\max\{{3+\lceil \log_2(\nu-1) \rceil}, 1+\nu,k_\mathrm{div}+2\}$ for the maximal depth of the certificates.
 In summary, the total depth of the scheme is given by $\max\{5+k_{\text{div}}+k_{\text{inv}}, 2+\nu\}$ assuming $\nu \geq 3$.
\setlength{\tabcolsep}{1.5mm}
\begin{table}[h]
    \centering
       \caption{Required multiplicative depth per block.}
    \label{tab:depthPerStage}
    \begin{tabular}{lll}
    \toprule
      Block / stage   &  Depth &   Source\\
      \midrule
      Preprocessing   & 1 & $\ct(M^\top M)$, $\ct(\mu)$, and $\ct(w_0)$ (parallel)\\
      \midrule
      Division & $1+k_{\text{div}}$ & Initialization and iterations \\
      Inversion & $2+k_{\text{inv}}$ & $\ct(\alpha)$, initialization, and iterations\\ 
      Least squares & $1$ & Multiplication $\ct(W_{k_\mathrm{inv}})\ct(V)$ \\
      \midrule
      Certificates & $1+\nu$ & Scaling, determinant, and multiplication\\
      \bottomrule
    \end{tabular}
\end{table}

In the numerical case study below, we have $\nu \in \{4,6,8\}$ and consider $k_{\mathrm{div}}=5$ and $k_{\mathrm{inv}}=12$. Hence, we need to setup the cryptosystem such that a multiplicative depth of $23$ is supported (if bootstrapping should be avoided). We do so by choosing the decryption modulus $Q_0=2^{60}$ and the rescaling factor $s=2^{30}$ leading to a ciphertext modulus of $Q_c\approx4.74 \times 10^{250}$. By setting a ring dimension of $N_r=2^{15}$ we achieve a security of approximately $133\;$bits according to the LWE estimator~\cite{albrecht2018estimate}, which satisfies the well-established $128$-bit security standard~\cite{albrecht2021homomorphic}. 
\section{Numerical case study}
\label{sec:Numerical}

\subsection{Setup}

To evaluate the performance of our scheme, we solve the three identification tasks from Section~\ref{subsec:sysID} for synthetic data. 
More precisely, we assume the system dynamics are specified by the SISO transfer function (TF)
\begin{equation}
  \label{eq:exemplary_system}
    G(z)=\frac{z^2+0.5z+2}{z^3+0.5z^2+0.25z+0.5}  
\end{equation}
with $n=3$. We then apply a random input signal of length $L=20$ with samples drawn from a standard normal distribution. The corresponding output signal is subject to normally distributed measurement noise with zero mean and standard deviation $10^{-3}$. For the identification of the state space model (SSM), where full state measurements are assumed, we consider the controllable canonical form of~\eqref{eq:exemplary_system}. For the multi-step predictor (MSP), we consider~$N=2$.

Now, regarding the parameters of our scheme, we assume the client aims for an error bound  $\epsilon=10^{-3}$ in \eqref{eq:desiredAccuracy}. Furthermore, we fix $k_{\mathrm{div}}=5$ and $k_{\mathrm{inv}}=12$ in all tasks for better comparability (as a constant multiplicative depth results).
The parameters $q$ and $\tau$ are chosen as $1$ (as suggested) and $1.999$, respectively. The remaining parameter $p$ is selected as the largest value for which \eqref{eq:epsilonCondition} holds in each task. Taking into account, that the triple $(l,\nu,r)$ reads $(17,6,1)$, $(19,4,3)$, and $(16,8,2)$ for the three tasks TF, SSM, and MSP, respectively, $p$ takes the value $0.997$ for all.

\subsection{Results}

\begin{figure}
    \centering
    \resizebox{\columnwidth}{!}{
    \includegraphics{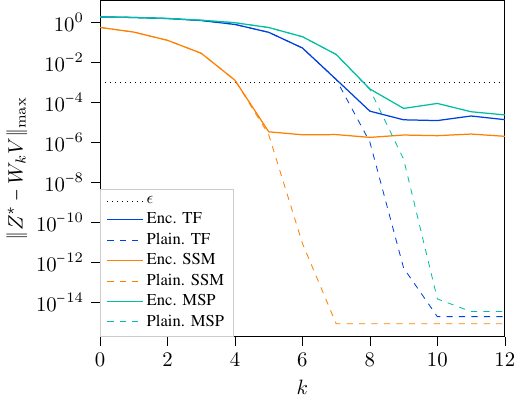}
    }
    \caption{Evolution of the approximation errors for the three identification tasks depending on the inversion iteration $k$. For each task, intermediate results of the encrypted implementation (solid) are compared with the plaintext counterpart (dashed).}
    \label{fig:Parameter_estimation_error}
\end{figure}

As apparent from Figure~\ref{fig:Parameter_estimation_error}, the identification can be carried out successfully for each task. In fact, the specified error bound is always satisfied.
Remarkably, this is also confirmed by the certificates for both the TF and SSM task. However, for the MSP task, the first certificate is invalid (as it is only sufficient but not necessary for the error bound to hold).
Figure~\ref{fig:Parameter_estimation_error} also confirms the quadratic convergence of the iterative matrix inversion underlying our approach. However, due to quantization and approximation errors naturally appearing in HE schemes, convergence of the encrypted scheme stops at some point.
In fact, the smallest error achieved by the encrypted system identification is significantly larger than for the plaintext counterpart. 
Nevertheless, for the considered case study, the achieved errors of order $10^{-5}$ are satisfactory when taking into account
 that $\|Z^\ast-Z_{\text{true}}\|_{\max}$ is of the order $10^{-4}$ in each case, where $Z_{\text{true}}$ reflects the true model parameters determined by~\eqref{eq:exemplary_system}.  
 Finally, we briefly comment on the required computation time for a single identification, which is on the order of $10$ minutes on a standard computer (with an AMD Ryzen 7 5800H CPU using 8GB RAM). Note, in this context, that the provided service is typically not real-time critical and that we did not consider any code optimization yet.

\section{Conclusions and Outlook}
\label{sec:Conclusions}

We presented an encrypted system identification as-a-service enabled by a reliable encrypted solution to least squares problems of the form~\eqref{eq:leastSquaresProb}.
The encrypted solution builds on encryption-friendly iterative algorithms to approximate the pseudo-inverse $M^\dagger$ central to the solution~\eqref{eq:bestFit}.
While encrypted versions of Newton-Schulz iterations~\eqref{eq:NewtonSchulzIter} or Goldschmidt-style iterations~\eqref{eq:GoldschmidtIter} have been used before for other applications, our contribution stands out for  
reliable initializations as well as certificates for the achieved accuracy without compromising the privacy of data provided by the client. The certificates can be prepared concurrently with the actual computations (see Fig.~\ref{fig:algorithm-overview-high-level}) leveraging the the computational capabilities of the service provider.
The effectiveness of the scheme has been illustrated by three numerical examples, where the certificates approved the desired accuracy in two cases. 

Future research directions are twofold. First, various improvements of the proposed certificates are promising.
This may include tighter bounds or more efficient implementations (e.g., w.r.t.~determinants).
Second, we here neglected the effect of quantization effects, naturally arising in HE, for our theoretical investigations. Formally including these effects in our approach would further increase its value.


\end{document}